\documentclass [12pt]{article}

\usepackage{amsmath,amsthm,amscd,amssymb}
\usepackage[polutonikogreek, english]{babel}
\usepackage[latin1]{inputenc}
\usepackage[linesnumbered]{algorithm2e}
\usepackage{placeins}
\usepackage{xcolor}
\usepackage[T1]{fontenc}
\usepackage[linesnumbered]{algorithm2e}

\usepackage[small, centerlast, it]{caption}
\usepackage{epsfig}
\usepackage[titles]{tocloft}
\usepackage[latin1]{inputenc}
\usepackage{verbatim} 
\usepackage[active]{srcltx} 
\usepackage{fancyhdr}
\usepackage{caption}

\def\sp{\hskip -5pt}

\def\b1{{1\!\!1}}

\def\sH{{\mathsf H}}

\def\bC{{\mathbb C}}           

\def\bI{{\mathbb I}}

\def\bR{{\mathbb R}}

\def\bx{{\textbf x}}

\def\beq{\begin{eqnarray}}
\def\eeq{\end{eqnarray}}

\SetKwInput{KwMode}{Mode}
\SetKwInput{KwParameters}{Parameters}

\def\by{\textbf{y}}

\def\p{\parallel}

\usepackage{sectsty}
\sectionfont{\normalsize}
\subsectionfont{\normalsize\normalfont\itshape}
\newtheoremstyle{thm}
{12pt}
{12pt}
{\itshape}
{}
{\itshape\bfseries}
{}
{1em}
{}

\theoremstyle{thm}

\usepackage{xy}
\xyoption{matrix}
\xyoption{frame}
\xyoption{arrow}
\xyoption{arc}

\usepackage{ifpdf}
\ifpdf
\else
\PackageWarningNoLine{Qcircuit}{Qcircuit is loading in Postscript mode.  The Xy-pic options ps and dvips will be loaded.  If you wish to use other Postscript drivers for Xy-pic, you must modify the code in Qcircuit.tex}
\xyoption{ps}
\xyoption{dvips}
\fi

\entrymodifiers={!C\entrybox}

\newcommand{\bra}[1]{{\left\langle{#1}\right\vert}}
\newcommand{\ket}[1]{{\left\vert{#1}\right\rangle}}
\newcommand{\qw}[1][-1]{\ar @{-} [0,#1]}
\newcommand{\qwx}[1][-1]{\ar @{-} [#1,0]}

\newcommand{\gate}[1]{*+<0.5em>{#1} \POS ="i","i"+UR;"i"+UL **\dir{-};"i"+DL **\dir{-};"i"+DR **\dir{-};"i"+UR **\dir{-},"i" \qw}
\newcommand{\meter}{*=<2.0em,1.5em>{\xy ="j","j"-<.778em,.322em>;{"j"+<.778em,-.322em> \ellipse ur,_{}},"j"-<0em,.4em>;p+<.5em,.9em> **\dir{-},"j"+<2.2em,2.2em>*{},"j"-<2.2em,2.2em>*{} \endxy} \POS ="i","i"+UR;"i"+UL **\dir{-};"i"+DL **\dir{-};"i"+DR **\dir{-};"i"+UR **\dir{-},"i" \qw}

\newcommand{\control}{*!<0em,.095em>-=-<.2em>{\bullet}}

\newcommand{\ctrl}[1]{\control \qwx[#1] \qw}

\newcommand{\targ}{*+<.04em,.04em>{\xy ="i","i"-<.50em,0em>;"i"+<.50em,0em> **\dir{-}, "i"-<0em,.47em>;"i"+<0em,.47em> **\dir{-},"i"*\xycircle<.6em>{} \endxy} \qw}
\newcommand{\qswap}{*=<0em>{\times} \qw}
\newcommand{\Qcircuit}{\xymatrix @*=<0em>}

\begin{document}

\begin{center}
{ \large \bfseries A quantum binary classifier based on cosine similarity}

\sp

{\sc {Davide Pastorello$^a$} and {Enrico Blanzieri}$^b$}

\bigskip

{
Department of Information Engineering and Computer Science, University of Trento, \\via Sommarive 9, 38123 Povo (Trento), Italy\\ ~~ $^a$ d.pastorello@unitn.it, $^b$ enrico.blanzieri@unitn.it\\[10pt]
}

\end{center}

{\small \section*{Abstract}  We introduce the quantum implementation of a binary classifier based on cosine similarity between data vectors. The proposed quantum algorithm evaluates the classifier on a set of data vectors with time complexity that is logarithmic in the product of the set cardinality and the dimension of the vectors. It is based just on a suitable state preparation like the retrieval from a QRAM, a SWAP test circuit (two Hadamard gates and one Fredkin gate), and a measurement process on a single qubit. Furthermore we present a simple implementation of the considered classifier on the IBM quantum processor ibmq\_16\_melbourne. Finally we describe the combination of the classifier with the quantum version of a $K$-nearest neighbors algorithm within a hybrid quantum-classical structure.}

\vspace{1cm}

\noindent
{\small \textbf{Keywords}: Quantum machine learning; binary classification; cosine similarity.}

\noindent
\section{Introduction}
Quantum machine learning is a rapidly emerging research area where quantum computing techniques are applied to machine learning tasks in order to pursue a computational advantage, in terms of time and space, within the present context of ever-growing amounts of data to manage. Within the supervised learning scenario, some interesting quantum classification algorithms have been proposed in recent years \cite{superv}. In particular in \cite{QSVM}, the authors present the quantum version of a support vector machine. The quantum SVM is based on advanced quantum subroutines like the \emph{quantum phase estimation} and the \emph{exponentiation of density matrices} that make its realization very hard with the available prototypes of quantum computers. However, there are some recent proposals of quantum classifiers assuming fewer quantum resources, like  a distance-based quantum classifier \cite{binaryclass} and a quantum implementation of the $K$-nearest neighbors algorithm \cite{QKNN}, that may represent more feasible solutions of quantum machine learning. In this paper, we introduce a quantum algorithm for binary classification that implements a model based on cosine similarity to predict the label of a new data instance. The main idea can be summarized as follows: The unclassified instance is put in the quantum superposition of the two possible classifications and then it is compared with all the training points with a single operation by quantum parallelism. 

The quantum classifier can be applied to a set of sample vectors, and
the set does not, in general, coincide withe whole training set. Given
a training set, the choice of the set can be made in different ways,
for example it could be the set of support vectors of a support vector
machine or it can be the neighborhood of the sample under a suitable
metric in an instance-based learning paradigm. Here we propose that
the classifier can be used in conjunction with a $K$-NN quantum
implementation that selects the set of neighbors given a
sample to classify.

In Section 2, we introduce the quantum algorithm for binary classification that assigns a label to an unclassified data vector evaluating the cosine similarity with all the training vectors. If $N$ is the number of training points and $d$ is the dimension of the feature space where data are represented, the proposed quantum algorithm classifies a new instance in time $O(\log (Nd))$ whereas the classical implementation of the same model presents a time complexity of $O(Nd)$. However we do not define a quantum model that can be trained but an efficient quantum algorithm to implement the binary classification according to a considered cosine similarity-based model. In Section 3, we introduce a low-sized example and perform a test on the IBM quantum processor ibmq\_16\_melbourne to check the correct classification of a data vector. In Section 4 we develop the idea of the combined use of the our
classifier with a quantum
implementation of $K$-NN. In Section 5 we draw the conclusions.

\sp

\section{The proposed quantum binary classifier}

Let $X=\{\bx_i, y_i\}_{i=0,...,N-1}$, with $\bx_{i}\in\bR^d$ and $y_i\in\{-1,1\}$ $\forall i\in\{0,...,N-1\}$, be a training set of $N$ data instances with two-valued labels that are represented in a real feature space of dimension $d$. Let $\bx\in\bR^d$ be a new data instance to be classified as either $1$ or $-1$. 
The classification model that we consider for the quantum implementation is defined as follows:
\beq\label{model}
y(\bx):=\mbox{sgn}\left(\sum_{i=0}^{N-1} y_i\cos (\bx_i, \bx)\right),
\eeq
where \emph{cosine similarity} is defined by:
\beq\label{cos}
\cos(\bx,\by):=\frac{\bx\cdot\by}{\p\bx\p \p\by\p}\qquad \bx,\by\in\bR^d.
\eeq 
A typical example where cosine similarity is adopted for classification and clustering is text analysis \cite{Hornik, cluster, tweets}. Furthermore, in the case of normalized data vectors, (\ref{cos}) reduces to the dot product and it is directly related to the Euclidean distance by $\p \bx-\by\p=\sqrt{2(1-\bx\cdot\by)}$. In the model (\ref{model}), any training vector contributes to the prediction of the new label and such a contribution is weighted by the cosine similarity with the new instance.

On one hand, the classical calculation of the new label according to (\ref{model}) requires $N$ cosine similarities each computed in time $O(d)$ then the overall time complexity is $O(Nd)$. On the other hand, the quantum implementation of the model (\ref{model}) that we introduce in the following is based on the encoding of data vectors into amplitudes of a coherent superposition of quantum states (amplitude encoding), on a suitable state preparation, and on the so-called SWAP test \cite{swap}.
 Assuming $d=2^n$ without loss of generality, the data vector $\bx_i\in\bR^d$ is encoded in the amplitudes of a quantum state of $n$ qubits:
\beq\label{amplitude}
\ket{\bx_i}=\frac{1}{\p\bx_i\p}\sum_{j=0}^{d-1} x_{ij}\ket j\in\sH_n,
\eeq
where $\ket j$ is an orthonormal basis of the $n$-qubit Hilbert space $\sH_n\simeq(\bC^2)^{\otimes n}$ and $x_{ij}$ is the $j$th component of $\bx_i$. Within the amplitude encoding we have the correspondence $\cos(\bx,\by)=\langle\bx|\by\rangle$. One of the key feature of quantum machine learning is that a quantum random access memory (QRAM) allows to retrieve data in parallel. Assuming that the real components $\{x_{ij}\}_{j=0,...,d-1}$ of $\bx_i$ are stored in an array of memory cells as floating point numbers and the norm $\p\bx_i\p$ is given separately (or we are considering normalized vectors), the retrieval of the state (\ref{amplitude}) can be done in $O(\log d)$ steps according to the \emph{bucket brigade architecture} of the QRAM introduced in \cite{Giovannetti}. Let us consider also a $\log N$-qubit register, with Hilbert space $\sH_{index}\simeq(\bC^2)^{\otimes \log N}$, to encode the indexes of training data vectors and construct the state:
\beq\label{X}
\ket X=\frac{1}{\sqrt N}\sum_{i=0}^{N-1}\ket i\ket{\bx_i}\ket{b_i}\in \sH_{index}\otimes\sH_n\otimes \sH_l,
\eeq
where $\sH_l$ is the Hilbert space of a single qubit used for encoding the values of the labels with $b_i=\frac{1-y_i}{2}\in\{0,1\}$, then $\ket {b_i}$ is eigenstate of the Pauli matrix $\sigma_z$ with eigenvalue $y_i$. The entangled state (\ref{X}) encodes the training set $X$ as a coherent superposition of its elements with respective labels, note that just one qubit is sufficient for the encoding of all the labels. Moreover, in the same registers we can construct the state:
\beq\label{psi}
\ket{\psi_\bx}=\frac{1}{\sqrt N}\sum_{i=0}^{N-1}\ket i \ket\bx\ket -\in \sH_{index}\otimes\sH_n\otimes \sH_l,
\eeq
where we have the label qubit in the state $\ket-=\frac{1}{\sqrt 2}(\ket 0-\ket 1)$, so the new data vector $\bx$ is represented in a quantum superposition of the two possible classifications.
Now consider an ancillary qubit, called qubit $a$, and let us prepare the state:
\beq\label{Phi}
\frac{1}{\sqrt 2} \left(\ket X \ket 0 +\ket{\psi_\bx}\ket 1\right)\in\sH_{index}\otimes\sH_n\otimes\sH_l\otimes\sH_a,
\eeq
that can be retrieved from the QRAM in time $O(\log(Nd))$. 
Now we perform the SWAP test (circuit (\ref{swap})) between a second ancillary qubit, called qubit $b$, prepared in $\ket +=\frac{1}{\sqrt 2}(\ket 0+\ket 1)$ and the qubit $a$, moreover we need another qubit, say $c$,  prepared in $\ket 0$ to control the Fredkin gate (i.e. the controlled swap gate):
\beq\label{swap}
\Qcircuit @C=1.0em @R=2.0em {
c\qquad&\gate{H}& \ctrl{1} & \gate{H}& \meter\\
b\qquad&\qw& \qswap & \qw& \qw\\
a\qquad&\qw& \qswap\qwx & \qw &\qw
} 
\eeq

\vspace{0.5cm}

A straightforward calculation\footnote{Appendix A for details.} 
 shows that after the action of (\ref{swap}) the probability to obtain the outcome 1 measuring the qubit $c$ is:
\beq\label{P}
\mathbb P(1)=\frac{1}{4}(1-\langle X|\psi_\bx\rangle),
\eeq
that is directly related to (\ref{model}), in fact we have:
\beq
\langle X|\psi_\bx\rangle=\frac{1}{N}\sum_{i,k=0}^{N-1}\langle i |k\rangle\langle\bx_i|\bx\rangle\langle b_i|-\rangle=
\frac{1}{N\sqrt 2}\sum_{i=0}^{N-1}\langle\bx_i|\bx\rangle(\langle b_i|0\rangle-\langle b_i|1\rangle)
\eeq
$$
=\frac{1}{N\sqrt 2}\sum_{i=0}^{N-1} y_i\cos(\bx_i,\bx),\hspace{5.6cm}
$$
where we have simply used the identities $\langle i| k\rangle=\delta_{ik}$ and $\langle b_i|0\rangle-\langle b_i|1\rangle=1-2b_i=y_i$ for any $i=0,...,N-1$. Therefore the probability $\mathbb P(1)$ is related to the prediction of the label of $\bx$, according to the model (\ref{model}), by means of:
\beq
y(\bx)=\mbox{sgn}\left[1-4\,\mathbb P(1)\right].
\eeq

After the retrieval of the state (\ref{Phi}) from a QRAM in time $O(\log (Nd))$, an appropriate SWAP test, performed in constant time, is sufficient to classify the new insatnce $\bx$ according to the model (\ref{model}) with an exponential speedup w.r.t. the classical calculation. However the procedure \emph{preparation+test} must be repeated several times for sampling the qubit $c$ to estimate $\mathbb P(1)$ as the success probability of a Bernoulli trial, so an estimation within an error $\epsilon$ requires a number of repetitions growing as $O(\epsilon^{-2})$ as provided by the binomial proportion confidence interval \cite{binomial}. Thus the overall time complexity of Algorithm \ref{qcos} is $O(\epsilon^{-2}\log{(Nd)})$.

\begin{algorithm}[ht!]
\footnotesize

\vspace{0cm}

\KwIn{training set $X=\{\bx_i,y_i\}_{i=0,...,N-1}$, unclassified instance $\bx$.} 
\KwResult{label $y$ of $\bx$.} 
\Repeat{desired accuracy on the estimation of $\mathbb P(1)$}{
initialize the register $\sH_{index}\otimes\sH_n\otimes \sH_l$ and an ancillary qubit $a$ in the state (\ref{Phi})\; 
initialize a qubit $b$ in the state $\ket -$\;
 perform the SWAP test on $a$ and $b$ with control qubit $c$ prepared in $\ket 0$; \emph{\% circuit (\ref{swap})}\\
measure qubit $c$\;}
Estimate the probability $\mathbb P(1)$ as the relative frequency $\hat{\mathbb P}$ of the outcome 1\;
\eIf{$\hat{\mathbb P}> 0.25$}{\Return $y=-1$}{\Return $y=1$}
\vspace{0.5cm}

\caption{\it Quantum implementation of the model (\ref{model}).}
\label{qcos}
\end{algorithm}

 \FloatBarrier

\section{Run on the IBM quantum processor}

Let us provide an example of implementation of the introduced quantum binary classifier for $N=2$ and $d=2$. Consider a training set of two-dimensional data instances given by $X=\{(\bx_0,y_0),(\bx_1,y_1)\}$ where $\bx_0=(1,0)$, $y_0=1$ and $\bx_1=(0.718, 0.696)$, $y_1=-1$. Let $\bx=(0.884,0.468)$ be the unlabelled data instance that we have to classify. In this simple example with normalized data vectors and only two training points, the model (\ref{model}) predicts the label of $\bx$ as a \emph{nearest neighbor} then it returns $y=-1$.

  In order to run Algorithm 1, according to (\ref{X}) and (\ref{psi}) we prepare the state:
\beq\label{XX}
\ket X=\frac{1}{\sqrt 2} (\ket 0\ket 0 \ket 0+\ket 1 \ket{\bx_1} \ket 1),
\eeq\label{psipsi}
and the state:
\beq
\ket{\psi_\bx}=\ket + \ket{\bx} \ket-.
\eeq
We run the algorithm on a IBM prototype quantum processor that is available within the cloud-based quantum computing service of \emph{IBM Quantum}. The machine is realized by superconducting transmon qubits located in a cryogenic refrigerator at the Thomas J. Watson Research Center, New York \cite{IBM}. Let us consider the quantum circuit for the construction of the state (\ref{XX}):
\beq
\Qcircuit @C=0.5em @R=1.5em {
\ket 0 \qquad&\gate{H}& \ctrl{1} & \ctrl{2}&\qw\\
\ket 0\qquad&\qw& \gate{RY_{(0.49 \pi)}} & \qw&\qw\\
\ket 0\qquad&\qw& \qw &  \targ &\qw
} 
\eeq

\noindent
where $RY_{(0.49 \pi)}$ is a rotation of $0.49 \pi$ around the $y$-axis of the Bloch sphere mapping $\ket 0$ into $\ket{\bx_1}=0.718\ket 0+0.696\ket 1$.
The circuit to construct the state (\ref{psipsi}) is simply:

\beq
\Qcircuit @C=0.4em @R=1.4em {
\ket 0 \qquad&\gate{H}& \qw\\
\ket 0\qquad& \gate{RY_{(0.31\pi)}} & \qw\\
\ket 0\qquad& \targ &\gate{H} &\qw
} 
\eeq

\sp\sp\sp

\noindent
where the gate $RY_{(0.31\pi)}$ rotates $\ket 0$ into $\ket\bx=0.884\ket 0+0.468\ket 1$ and we have a bit-flip followed by a Hadamard gate to prepare the label qubit in $\ket -$.

In Fig. 1, there is the circuit, represented in the \emph{IBM Quantum Composer},  implementing Algorithm 1 for the example just introduced. The qubits $q_0, q_1, q_2$ are the ancillas used for the SWAP test, $q_3$ is the 1-qubit index register, $q_4$ is an additional ancillary qubit necessary to implement the controls of the gate $RY_{(0.49 \pi)}$, $q_5$ is the 1-qubit register for the amplitude encoding of data and $q_6$ is the qubit encoding the labels.

\vspace{0.5cm}

\begin{center}

\includegraphics[width=14cm]{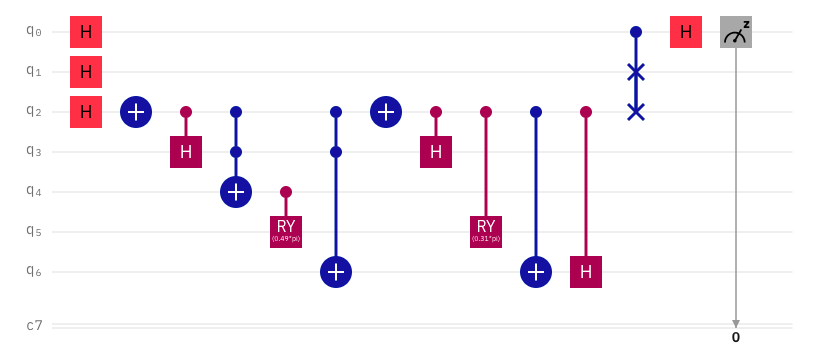}

\end{center}

\vspace{-0.3cm}

\noindent
{\small \textbf{Fig.1}. Quantum circuit implementing lines 1-7 of Algorithm 1 in the IBM Quantum Composer where $N=2$, $d=2$, the training set is $\{((1,0),1) \,,\, ((0.718, 0.696), -1)\}$ and the unclassified data vector is $\bx=(0.884,0.468)$.  }

\sp\sp\sp\sp\sp

We have run the algorithm on the IBM quantum processor \emph{ibmq\_16\_melbourne} performing 1024 shots for sampling the qubit $q_0$ (corresponding to the qubit $c$ of our general algorithm description). The obtained statistic is shown in Fig. 2, the relative frequency of the outcome 1, used to estimate the probability $\mathbb P(1)$, is $\hat{\mathbb P}=447/1024>0.25$ then the label assigned to $\bx=(0.884,0.468)$ is $y=-1$ as expected.

\vspace{-0cm}

\begin{center}
\includegraphics[width=6.5cm]{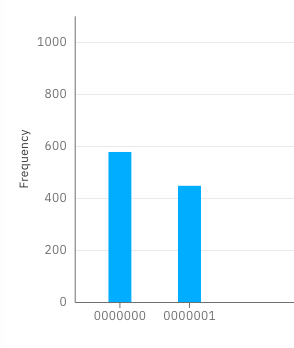}
\end{center}
\vspace{-1.1cm}
\begin{center}
{\small \qquad\quad\textbf{Fig.2}. Output statistic of the circuit reported in Fig.1.}
\end{center}

\section{Combination with the quantum K-NN}

The proposed binary classifier can be combined with a quantum version of the \emph{K-nearest neighbors algorithm} presented in \cite{QKNN}. In fact, given the training set $X=\{\bx_i, y_i\}_{i=0,...,N-1}$ and the unclassified instance $\bx$ we can assume to select the $K$ closest training points to $\bx$ (w.r.t. cosine similarity) and then performing the quantum binary classification described above. So we do not consider the whole training set for the label assignment but just the training points in a neighborhood of the new data vector. 
We focus on vectors with positive entries, this is not a sever restriction if we take into account that data vectors are so in text analysis that is the most relevant framework where cosine similarity is used for classification and clustering.

The selection of the $K$ nearest neighbors can be done implementing a SWAP test as well. However the \emph{controlled swap} is not a standard Fredkin gate that conditionally swaps only two qubits but we must swap two $n$-qubit registers. It is straightforward to show that the controlled swap of two $n$-qubit registers can be implemented with $n$ Fredkin gates controlled by the same qubit. 

Let us consider two $n$-qubit register (i.e. two copies of the main register where data are encoded), the index register, and an ancillary qubit prepared in the state:
\beq\label{knn1}
\frac{1}{\sqrt N}\sum_{i=0}^{N-1} \ket i\ket{\bx_i}\, \ket{\bx}\,\ket 0\in \sH_{index}\otimes\sH_n\otimes\sH_n\otimes\sH_a.
\eeq
In (\ref{knn1}), we have the superposition of the training points and the new data instance stored in two different registers, instead in the classification algorithm described above these data are superposed in the same register.
Now perform the SWAP test on the two $n$-qubit registers controlled by the ancillary qubit obtaining:
\beq\label{Psii}
\ket{\Psi}=\frac{1}{2\sqrt N} \sum_{i=0}^{N-1} \ket i \left[(\ket{\bx_i}\ket{\bx}+\ket{\bx}\ket{\bx_i})\ket 0+(\ket{\bx_i}\ket{\bx}-\ket{\bx}\ket{\bx_i})\ket 1\right].
\eeq
The probability to obtain outcome $\alpha\in\{0,1\}$ by a measurement process on the ancillary qubit is:
\beq\label{P1}
\mathbb P(\alpha)=\frac{1}{2}+(-1)^{\alpha}\frac{1}{2N}\sum_{i=0}^{N-1}|\langle\bx|\bx_i\rangle|^2,
\eeq
and the corresponding post-measurement state of the registers is:
\beq\label{psi0}
\ket{\Psi_\alpha}=\frac{\sum_{i=0}^{N-1} \ket i(\ket{\bx_i}\ket{\bx}+(-1)^{\alpha}\ket{\bx}\ket {\bx_i})}{\sqrt {2\left(N+(-1)^\alpha\sum_{i=0}^{N-1} |\langle\bx|\bx_i\rangle|^2\right)}}\, \ket \alpha.
\eeq
Once obtained the outcome $\alpha$ measuring the ancillary qubit, we can perform a measurement on the index register, the probability\footnote{The probabilities (\ref{P1}) and (\ref{P2}) and the state (\ref{psi0}) are explicitly calculated in Appendix B.} to obtain the outcome $i$ is:
\beq\label{P2}
\mathbb P(i|\alpha)=\frac{ 1+(-1)^\alpha|\langle\bx|\bx_i\rangle|^2}{ {N+(-1)^\alpha\sum_{i=0}^{N-1} |\langle\bx|\bx_i\rangle|^2}}.
\eeq
We have:
\beq\label{knn2}
\mathbb P(i|0)-\mathbb P(i|1)=\frac{2(|\langle\bx|\bx_i\rangle|^2- C)}{N(1-{C^2})},
\eeq
where $C=\frac{1}{N}\sum_{i}|\langle\bx|\bx_i\rangle|^2$ is a constant, so (\ref{knn2}) is proportional to the square cosine similarity between $\bx_i$ and $\bx$ that is the \emph{fidelity} between the quantum states $\ket\bx$ and $\ket{\bx_i}$. Since we are considering only vectors with positive entries, sampling the index register we obtain the most likely indexes corresponding to the closest vectors to $\bx$. Then we can run the binary classifier defined above without evaluating any training point but only those in a neighborhood of the new data instance. 

Algorithm \ref{hybrid} is a hybrid quantum-classical algorithm for binary classification providing the quantum selection of the $K$ closest training points to the unlabelled data vector and the following assignment of the label running Algorithm \ref{qcos}. In the algorithm there are classical operations to read the output of the $K$-NN and initialize the quantum classification routine in a neighborhood of the unclassified vector.

\begin{algorithm}[ht!]
\footnotesize

\vspace{0cm}

\KwIn{training set $X=\{\bx_i,y_i\}_{i=0,...,N-1}$, unclassified instance $\bx$.} 
\KwResult{label $y$ of $\bx$.} 

initialize the register $\sH_{index}\otimes\sH_n\otimes\sH_n$ and an ancillary qubit $a$ in the state (\ref{knn1})\; 
perform the SWAP test on the $n$-qubit registers controlled by the qubit $a$\;
measure the qubit $a$\;
{compute (\ref{knn2}) sampling the index register\;
select the $K$ most likely indexes $\{i_1,...,i_K\}$\;
$X\gets \{\bx_{i_k}, y_{i_k}\}_{k=1,...,K}$\;
$y\gets Algorithm\,\ref{qcos}(X,\bx)$\;
\Return $y$\;}
\vspace{0.5cm}

\caption{\it Hybrid quantum-classical algorithm for binary classification.}
\label{hybrid}
\end{algorithm}

 \FloatBarrier

\section{Conclusions}

In this paper we have presented a quantum implementation of an elementary binary classifier based on cosine similarity with an exponential speedup w.r.t. to any classical implementation of the same model. Then we have tested the proposed quantum algorithm on a IBM quantum processor to show that it can be easily run on available quantum machines. The algorithm shares some characteristics with known quantum machine learning algorithms based on the SWAP test \cite{Lloyd, QSVM, QKNN}. The peculiarity of our proposal is the quantum calculation of the cosine similarity between the unclassified vector and all the training vectors weighted with the labels without parameters that must be learned in a training phase. A natural application framework of this classification model may be text analysis where documents are typically compared by means of cosine similarity. 

We also provided and example of use of our classifier in conjunction
with a quantum implementation of the $K$-nearest neighbors. The
classifier is applied to the $K$ vectors outputted by the $K$-NN. Although
the combination is hybrid quantum-classical in nature and further work
is needed in order to integrate the two in a completely quantum
version, it illustrates the potentiality of our proposal to be a
component of novel quantum machine learning algorithms.

The exponential speedup of the quantum classifier is due to the efficient preparation of quantum states in logarithmic time and to the classification itself performed in constant time (that depends on the desired accuracy). However the efficient retrieval is a common bottleneck in quantum machine learning as the QRAM architecture is extremely expansive in terms of space resources. Thus we conclude that new effective procedures for representing classical data into quantum hardware are a crucial direction of investigation for quantum machine learning. 
Moreover a goal of the pivotal proposal of this paper is the suggestion of a general approach in devising quantum machine learning algorithms that should have the requirement of being easily and efficiently implemented on the available or near-term quantum machines of the NISQ era.   

\section*{Acknowledgements} 

The work of D.P. is supported by a grant of the Q@TN consortium. 


\sp

\section*{Appendix A: Derivation of (\ref{P})}

Consider the circuit (\ref{swap}) for the SWAP test: If the input state is $\ket 0_c\ket\varphi_b\ket\psi_a$, we have the following processing:
\beq
\ket 0_c\ket\varphi_b\ket\psi_a\xrightarrow[]{H\otimes\bI_{b}\otimes\bI_a}\frac{1}{\sqrt 2}(\ket{0}_c+\ket 1_c)\ket\varphi_b\ket\psi_a\xrightarrow[]{Fredkin\,\, gate}\frac{1}{\sqrt 2}(\ket{0\,\varphi\,\psi}_{cba}+\ket {1\,\psi\,\varphi}_{cba})
\eeq
$$
\xrightarrow[]{H\otimes\bI_{b}\otimes\bI_a}\frac{1}{2}\ket 0_c(\ket{\varphi\,\psi}_{ba}+\ket{\psi\,\varphi}_{ba})+\frac{1}{2}\ket 1_c(\ket{\varphi\,\psi}_{ba}-\ket{\psi\,\varphi}_{ba}),\qquad
$$
where $\bI_{a,b}$ are the identity operators acting on the qubits $a$ and $b$.
Therefore the probability of measuring 1 on the qubit $c$ is:
\beq
\mathbb P(1)= \frac{1}{4}(\bra{\varphi\,\psi}_{ba}-\bra{\psi\,\varphi}_{ba})(\ket{\varphi\,\psi}_{ba}-\ket{\psi\,\varphi}_{ba})=\frac{1}{2}(1-|\langle\varphi|\psi\rangle|^2),
\eeq
so the SWAP test is a useful and well-known tool to estimate $|\langle\varphi|\psi\rangle|^2$ given two unknown pure states $\ket\psi$ and $\ket\varphi$. 

In our case the qubit $b$ is in $\ket +$ and the qubit $a$ is entangled with the $n$-qubit register encoding the data as provided by the state (\ref{Phi}). Then the probability to obtain $1$ by a measurement on the qubit $c$ is $
\mathbb P(1)=\frac{1}{2}(1-\langle \Phi|\Phi\rangle)$
where $\Phi$ is the non-normalized vector obtained taking the inner product of $\ket +$ and the ancillary tensor factor of the state (\ref{Phi}):
\beq
\Phi=\frac{1}{\sqrt 2}(\langle +|0\rangle \ket X+\langle +|1\rangle\ket{\psi_{\bx}})=\frac{1}{2}(\ket X+\ket{\psi_{\bx}}).
\eeq
Therefore:
\beq
\mathbb P(1)=\frac{1}{2}(1-\langle \Phi|\Phi\rangle)=\frac{1}{2}\left[1-\frac{1}{4}(2+2\langle X|\psi_{\bx}\rangle)\right]=\frac{1}{4}(1-\langle X|\psi_{\bx}\rangle).
\eeq

\sp

\section*{Appendix B: Derivation of (\ref{P1}), (\ref{psi0}), (\ref{P2})}

The probability to obtain the outcome $0$ by a measurement process on the ancillary qubit in the state (\ref{Psii}) is:
\beq
\mathbb P(0)=\langle\Psi| P_0\Psi\rangle\quad\mbox{with}\quad P_0=\bI_{index,n}\otimes \ket 0\bra 0,
\eeq
where $\bI_{index, n}$ is the identity operator on the Hilbert space $\sH_{index}\otimes \sH_n^{\otimes 2}$ and $\ket 0\bra 0$ is the projector onto the 1-dimensional subspace of $\sH_a$ spanned by $\ket 0$, as provided by the postulates of quantum mechanics.
Therefore:
\beq
\mathbb P(0)=\frac{1}{4N}\sum_{i,l=0}^{N-1} \langle i|l\rangle (\bra{\bx_i}\bra\bx+\bra\bx\bra{\bx_i})(\ket{\bx_l}\ket\bx+\ket\bx\ket{\bx_l})= 
\eeq
$$\quad=\frac{1}{4N} \sum_{i=0}^{N-1} (2 + 2\langle\bx_i|\bx\rangle\langle\bx|\bx_i\rangle) 
=\frac{1}{2} +\frac{1}{2N} \sum_{i=0}^{N-1} |\langle\bx|\bx_i\rangle|^2. $$
\\
A similar calculation, considering the projector $P_1=\bI\otimes \ket 1\bra 1$, yields $\mathbb P(1)=\langle\Psi|P_1\Psi\rangle=\frac{1}{2} -\frac{1}{2N} \sum_{i=0}^{N-1} |\langle\bx_i|\bx\rangle|^2$. Thus we have (\ref{P1}).
\\
The post-measurement state (\ref{psi0}) is obtained normalizing the projected vector $P_\alpha\ket\Psi$.
\\
The probability (\ref{P2}) is given by:
\beq
\mathbb P(i|\alpha)=\langle\Psi_\alpha | Q_i \Psi_\alpha\rangle \quad\mbox{with}\quad Q_i=\ket i\bra i\otimes \bI_{n,a},
\eeq
where $\bI_{n,a}$ is the identity operator on the Hilbert space $\sH_n^{\otimes 2}\otimes \sH_a$ and $\ket i\bra i$ is the projector onto the subspace of $\sH_{index}$ spanned by $\ket i$. Therefore:
\beq
\mathbb P(i|\alpha)=\frac{ (\bra{\bx_i}\bra{\bx}+(-1)^{\alpha}\bra{\bx}\bra {\bx_i})(\ket{\bx_i}\ket{\bx}+(-1)^{\alpha}\ket{\bx}\ket {\bx_i})}{ 2({N+(-1)^\alpha\sum_{i=0}^{N-1} |\langle\bx|\bx_i\rangle|^2})}=
\eeq
$$\hspace{2.1cm}= \frac{2+2(-1)^\alpha \langle\bx_i|\bx\rangle\langle\bx|\bx_i\rangle}{ 2({N+(-1)^\alpha\sum_{i=0}^{N-1} |\langle\bx|\bx_i\rangle|^2})} =\frac{ 1+(-1)^\alpha|\langle\bx|\bx_i\rangle|^2}{ {N+(-1)^\alpha\sum_{i=0}^{N-1} |\langle\bx|\bx_i\rangle|^2}}. $$

\sp


\begin{thebibliography}{}

\bibitem[Al16]{cluster} S. Al-Ananzi, H. AlMahmoud, I. Al-Turaiki \emph{Finding similar documents using different clustering techniques} Procedia Computer Science 82 (2016) 

\bibitem[BG20]{QKNN} A. A. Basheer, S. K. Goyal \emph{Quantum k-nearest neighbor machine learning algorithm} preprint (2020) 

\bibitem[BCD01]{binomial} L. D. Brown, T. T. Cai, A. DasGupta \emph{Interval Estimation for a Binomial Proportion} Statist. Sci. 16 (2) 101 - 133 (2001)

\bibitem[Bu01]{swap}
H. Buhrman, R. Cleve, J. Watrous, R. de Wolf \emph{Quantum Fingerprinting}
Phys. Rev. Lett. 87, 167902 (2001)

\bibitem[DM01]{Dhillon} I. S. Dhillon, D. S. Moda \emph{Concept decomposition for large sparse text data using clustering} Machine Learning, 42, 143-175 (2001) 

\bibitem[FA17]{tweets} C. Focil-Arias, J. Ziiniga, G. Sidorov, I. Batyrshin, A. Gelbukh
\emph{A tweets classifier based on cosine similarity} CEUR Workshop Proceedings (2017)

\bibitem[Ho12]{Hornik} K. Hornik, I. Feinerer, M. Kober, C. Buchta \emph{Spherial $k$-means clustering} Journal of Statistical Software vol. 50, n. 10 (2012)

\bibitem[IBM21]{IBM} https://quantum-computing.ibm.com

\bibitem[GLM08]{Giovannetti} V. Giovannetti, S. Lloyd, L. Maccone \emph{Quantum random access memory} Phys. Rev. Lett. 100, 160501 (2008)

\bibitem[Ll13]{Lloyd} S. Lloyd, Mohseni, M., Rebentrost, P. \emph{Quantum algorithms for supervised and unsupervised machine learning}. Preprint at https://arxiv.org/abs/1307.0411 (2013)

\bibitem[Re14]{QSVM} P. Rebentrost, M. Mohseni, S. Lloyd \emph{Quantum support vector machine for big data classification} Phys. Rev. Lett. 113, 130503 (2014) 


\bibitem[Sc17]{binaryclass} M. Schuld et al. \emph{Implementing a distance-based classifier with a quantum interference circuit} EPL 119 60002 (2017)

\bibitem[SP18]{superv} M. Schuld, F. Petruccione \emph{Supervised learning with quantum computers} Springer International Publishing 2018


\end{thebibliography}
\end{document}